   \documentclass[prl,aps,twocolumn,showpacs]{revtex4}
                                                               \usepackage[dvips]{graphicx}
\usepackage{dcolumn}
                                                                           
\newcommand{\be}{\begin{equation}}
\newcommand{\ee}{\end{equation}}
\newcommand{\bea}{\begin{eqnarray}}
\newcommand{\eea}{\end{eqnarray}}
\newcommand{\nn}{\nonumber}

\begin{document}
\topmargin=-20mm
%      \Large

\title{Low Temperature Electronic Transport and  Electron
Transfer through Organic Macromolecules}
\author{{\bf Natalya Zimbovskaya}}
\address{Department of Physics
City College of CUNY, New York, NY 10031}

 \begin{abstract}
 It is shown that at low temperatures and moderate electron dephasing the electron transmission function reveales a structure containing information about donor/acceptor sites effectively participating in the electron transfer  processes and primary pathways of electrons tunneling through molecular bridges in macromolecules. This important information
can be obtained as a result of analysis of experimental low temperature current-voltage characteristics for chosen molecules.
 \end{abstract}


\date{\today}
\maketitle

%\begin{multicols}{2}
%\narrowtext

%\newpage

It has been estabilished that molecular electron transfer (ET) is essentially a combination of nuclear environment fluctuations and electron tunneling. Due to the large
distances between donor and acceptor, ET is mostly provided by 
intervening molecular bridges, giving rise to a set of intermediate states for the electron tunneling \cite{1}. The expression for the ET rate including both electronic and nuclear factors was first proposed by Marcus \cite{2,3,4}
and can be written as follows: 
     \begin{equation}
K_{ET} = K_{el} K_n \nu_n \, . 
    \label{e1}       \end{equation}
 Here, $ K_{el}$ is the electron transmission coefficient, $ K_n $ is the nuclear transition or Franck-Condon factor, and $ \nu_n $ is the effective nuclear vibration frequency.

The enormous size of biological molecules makes calculations of $ K_{el}$ extremely difficult \cite{5,6,7,8,9,10,11}. In this paper a simple approach is proposed which enables us to show that some intrinsic characteristics of the intramolecular ET such as pathways of tunneling electrons and donor/acceptor coupling to the bridge at different values of tunnel energy $ E $ can be obtained in experiments on low-temperature
electrical transport through the corresponding molecules. These data may be available under moderate electronic dephasing \cite{12,13,14}. A similar approach could also be developed to analyze characteristics of intermolecular ET processes in extended biosystems.

Existing theories of electronic transport through macromolecules are mostly based on a very simple model simulating both electrodes by semi-infinite tight-binding chains attached to the ends of the molecule which is also simulated as a single tight-binding chain of sites \cite{12,13,14,15,16,17}. Here we adopt a resembling model.
However, we take into account that donor and acceptor subsystems in realistic macromolecules are complex and include many sites providing effective coupling to the bridge, as well as to the attached electrodes. 

To simplify further calculations we assume coherent electron tunneling through the macromolecule to be the predominant transport mechanism. Correspondingly we treat the electron transport through the molecule as a sequence of tunnelings between potential wells. Each well represents one
atomic orbital of the donor-bridge-acceptor system. Within this approach, any atom is represented by a set of sites corresponding to its states. For further simplification we
assume than intraatomic hopping integrals are smaller that interatomic ones, therefore we consider electron transfer between different sitesas its tunneling between different atoms.

Assume that sites ''i'' are associated with those atoms of the bridge which have a strong coupling to the donor, and sites ``$j$'' are associated with atoms interacting with the acceptor.  Then we can treat any site ``$i$'' as the origin of a chain along which an electron moves from donor to acceptor, and any site ``$j$'' as the end of the chain. Thus, we have a set of chains (pathways) for an electron moving along the bridge. In further analysis, we concentrate on the case when the chains
weakly interact so we can consider them separately.
We simulate the donor/acceptor subsystems as sets of semi-infinite tight-binding homogeneous chains. Each chain is attached to a site of donor/acceptor which can be effectively coupled to the bridge. As before we assume that the chains do not interact with each other. The adopted model does not enable us to carry out a proper treatment of coupling of electrodes to the molecule at metal-molecule-metal junctions which is a nontrivial problem \cite{18,19,20,21}. However,  it seems reasonable to conjecture that in the molecules with complex donor and acceptor subsystems this process do not strongly depend on details of coupling of the electrodes to the donor/acceptor due to a comparatively large size and complicated structure of these subsystems.

In the following calculations we start from an effective tight-binding
Hamiltonian for a single chain included into the bridge:

  \begin{equation}
H_{eff} = H_0 + H_1 + H_{\cal D} + H_{\cal A}\ .
   \label{e2}
       \end{equation}
 Here first two terms describe the chain itself, and their matrix
elements between states $|k>$ and $|l>$ corresponding to the $k$th and
$l$th sites are given by:

 \begin{equation}
(H_0 )_{kl}= \alpha_k\delta_{kl} \  ; \qquad\qquad
(H_1 )_{kl}=  {\cal V}_{kl}\ ,
     \label{e3}
           \end{equation}
 where
${\cal V}_{kl}$= 0 when $k =
l$ and only states associated with valence electrons are considered.  The
diagonal matrix elements $\alpha_k$ are ionization energies of electrons
at sites $k$, while ${\cal V}_{kl} = {\cal V}_{lk}$ includes both direct
and exchange energy contributions for an electron to transfer between the
$k$th and $l$th sites.  
Remainig terms represent self-energy corrections
arising due to the coupling of the donor $(H_{\cal D})$ and acceptor
$(H_{\cal A})$ to the bridge:
  \bea %f4-5
  (H_{\cal D})_{ii} =
(\Sigma_{\cal D} )_{i}=\sum_m \frac{{\cal D}_{m i}^{\ 2}}
{E-\epsilon_m-\sigma_m}\ , &&
      \label{e4}
          \\ \nn \\ 
(H_{\cal D})_{jj} =
(\Sigma_{\cal A})_{j}=\sum_n \frac{{\cal A}_{jn}^{\
2}}{E- \epsilon_n - \sigma_n}\ . &&
  \label{e5}
        \eea
  Here, ${\cal D} _{mi}$ and ${\cal A}_{jn}$ are, respectively, coupling strengths between the $m$th donor site or the $n$th acceptor site and the $i$th or $j$th site of the bridge, and $\sigma_{m,n} = \frac{1}{2} \left\{ \theta_{m,n} - i \sqrt{4 \gamma_{m,n}^2 - \theta_{m,n}^2}\right\}$ are the self-energy corrections of the semi-infinite chains attached to
the corresponding sites \cite{12}.  The parameters $\theta_{m,n} = E - \epsilon_{m,n},$ where $ \epsilon_{m,n}$ and $\gamma_{m,n}$ are the ionization energies of electrons at the corresponding donor/acceptor sites, and the nearest-neighbor hopping integrals for the chains.  Summation in Eqs. (4) and (5) is carried out over all donor/acceptor sites coupled to the bridge.        
Due to the presence of the self-energy corrections the eigenvalues of the effective Hamiltonian [Eq. (2)] include imaginary parts which give broadening of the bridge energy levels $E_i.$  The energy levels are broadened further if include scattering processes in the bridge which are
not considered here.

%%% 8 -10
 An electric tunneling current $I$ flowing from donor to acceptor through the bridge in the presence of a small  voltage $V$ applied across the bridge has the form \cite{22}
   \begin{equation} %f6
I = \frac{e}{\pi \hbar}\int_{- \infty}^{ \infty}dE\ T(E) \lbrack f \left(E - {\mu}_1
\right)- f \left( E - {\mu}_2 \right) \rbrack\ ,
 \label{e6}
        \end{equation}
 where $f(E)$ is the Fermi function, and the chemical potentials ${\mu}_1$ and ${\mu}_2$ are determined by the equilibrium Fermi energy of the bridge $E_F$ and the effective voltage $V$ across the bridge \cite{16}, i.e.
  \[
{\mu}_1 = E_F + \left( 1 - \eta \right) eV;
\qquad  {\mu}_2 = E_F - \eta e V.
              \]
 Here, the parameter $ \eta $ characterizes how the voltage $ V $ is divided between the two ends of the bridge. The electron transmission function is given by the formula 
  \begin{equation} %f7
T(E) = 2 \sum_{i,j} \Delta_i |G_{ij}|^2 \Delta_j \ .
   \label{e7}
      \end{equation}

The summation in Eq. (7) is carried over the bridge states $|i>$ and  $ |j> $  therefore contributions from all possible
pathways are contained here. The quantities $ \Delta_{i,j} $ are imaginary parts of the self-energy corrections $ \sum_{\cal D,A}, $ and   $G_{ij}$ is the matrix element of the Green's function corresponding to $H_{{eff}} :$ 
  \begin{equation} %f8
{\cal G}_{ij}=\left<i\left|\left( E- H_0 -H_1 - H_{{\cal D}} - H_{{\cal A}} \right)^{-1} \right|  j \right>\ .
       \label{e8}
             \end{equation}
 We see that the dependence of the electron transmission function on energy is determined by the contributions from different donor/acceptor sites, as well as from the Green's function matrix elements corresponding to different chains included into the bridge subsystem.

 We can easily arrive at the first approximation for $ G_{ij} $ neglecting the broadening of the bridge energy levels. Taking into account only nearest neighbors (NN) and next nearest neighbor (NNN) couplings, we can introduce a new notation for the nonzero hopping integrals, i.e., $ {\cal V}_{k,k+1} = {\cal V}_{k-1,k} = \beta_k $ and $ {\cal V}_{k,k+2} = {\cal V}_{k-2,k} = \gamma_k .$
 Then we can  calculate the Green's function matrix elements for a given chain which makes a start at the site ''i'' and ends at the site $''j'',$  applying a diagram technique similar to that described in detail for the (NN) approximation \cite{23}. We present our chain of sites as a graph $\Gamma_{ij},$ using the following rules:

{(i)} Any site included in the chain corresponds to a 
graph vertex.

{(ii)} A term $E-\alpha_k$ corresponds to a loop attached to the
vertex ``$k$''.

{(iii)} A hopping integral ${\cal V}_{kl}$ corresponds to an edge 
originating at the vertex ``$k$'' and ending at the vertex ``$l$''.

As a result we obtain the graph shown below:

\vspace{12mm}
%%% 18
\begin{picture}(0,0)(-48,-30)

\put(-60,-45){\framebox(18,14){i}}  \put(-51,-21){\circle{20}}
\put(-23,-45){\framebox(18,14){i+1}}  \put(-14,-21){\circle{20}}
\put(15,-45){\framebox(18,14) {i+2}}   \put(24,-21){\circle{20}}
\put(81,-45){\framebox(18,14){k}}   \put(89,-21){\circle{20}}
\put(116,-45){\framebox(18,14){k+1}}   \put(125,-21){\circle{20}}
\put(154,-45){\framebox(18,14){k+1}}  \put(162,-21){\circle{20}}

\put(-40,-38){\vector(1,0){16}}  \put(-25,-38){\vector(-1,0){16}}
\put(-3,-38){\vector(1,0){17}}   \put(12,-38){\vector(-1,0){17}}
\put(65,-38){\vector(1,0){15}}   \put(48,-38){\vector(-1,0){15}}
\put(99,-38){\vector(1,0){17}}  \put(116,-38){\vector(-1,0){17}}
\put(136,-38){\vector(1,0){17}}   \put(152,-38){\vector(-1,0){17}}
\put(173,-38){\vector(1,0){16}}   

\put(-62,-8){$E-\alpha_i$}  \put(-36,-34){$\beta_{i}$}
\put(-30,-8){$E-\alpha_{i+1}$}  \put(-4,-34){$\beta_{i+1}$}
\put(10,-8){$E-\alpha_{i+2}$}   \put(34,-34){$\beta_{i+2}$}
\put(73,-8){$E-\alpha_{k}$}    \put(61,-34){$\beta_{k-1}$}
\put(107,-8){$E-\alpha_{k+1}$}    \put(103,-34){$\beta_k$}
\put(150,-8){$E-\alpha_{k+2}$}   \put(134,-34){$\beta_{k+1}$}

\put(173,-34){$\beta_{k+2}$}
\put(51,-40){$\cdots$}

\put(-17,-61){$\gamma_{j}$}
\put(6,-61){$\gamma_{j+1}$}
\put(32,-61){$\gamma_{j+2}$}
\put(62,-61){$\gamma_{k-2}$}
\put(90,-61){$\gamma_{k-1}$}
\put(122,-61){$\gamma_{k}$}
\put(145,-61){$\gamma_{k+1}$}
\put(172,-61){$\gamma_{k+2}$}

%\put(-15,-46){\oval(74,26)[b]}
\put(-6,-55){\line(-1,0){17}} \put(24,-55){\line(-1,0){9}}
\put(-6,-55){\vector(3,1){28}}
\put(-23,-55){\vector(-3,1){28}} 
\put(15,-55){\vector(-3,1){28}} \put(50,-55){\vector(-3,1){28}}

\put(135,-55){\line(-1,0){17}} \put(98,-55){\line(-1,0){9}}
\put(62,-55){\vector(3,1){28}}
\put(98,-55){\vector(3,1){28}}
\put(135,-55){\vector(3,1){28}}
\put(118,-55){\vector(-3,1){28}}
\put(153,-55){\vector(-3,1){28}} \put(153,-55){\line(1,0){9}} 
\put(192,-55){\vector(-3,1){28}}

\end{picture}
\vspace{21mm}

\begin{picture}(0,0)(-86,-30)

\put(-60,-45){\framebox(18,14){j--2}}  \put(-51,-21){\circle{20}}
\put(-21,-45){\framebox(18,14){j--1}}  \put(-12,-21){\circle{20}}
\put(17,-45){\framebox(18,14) {j}}   \put(26,-21){\circle{20}}
%\put(81,-45){\framebox(18,14){k}}   \put(89,-20){\circle{20}}
%\put(116,-45){\framebox(18,14){k+1}}   \put(125,-20){\circle{20}}
%\put(154,-45){\framebox(18,14){k+1}}  \put(162,-20){\circle{20}}

\put(-40,-38){\vector(1,0){18}}  \put(-25,-38){\vector(-1,0){16}}
\put(-2,-38){\vector(1,0){18}}   \put(14,-38){\vector(-1,0){17}}
\put(-78,-38){\vector(1,0){17}}   
%\put(65,-38){\vector(1,0){14}}  \put(47,-38){\vector(-1,0){14}}
%\put(99,-38){\vector(1,0){17}}  \put(116,-38){\vector(-1,0){17}}
%\put(136,-38){\vector(1,0){17}}   \put(152,-38){\vector(-1,0){17}}

\put(-71,-8){$E-\alpha_{j-2}$}  \put(-42,-34){$\beta_{j-2}$}
\put(-29,-8){$E-\alpha_{j-1}$}  \put(-3,-34){$\beta_{j-1}$}
\put(13,-8){$E-\alpha_{j}$}  \put(-81,-34){$\beta_{j-3}$}

\put(-92,-40){$\cdots$}

\put(-4,-55){\line(-1,0){18}} \put(-41,-55){\line(-1,0){9}}
\put(-4,-55){\vector(3,1){28}}
\put(-22,-55){\vector(-3,1){28}} 
\put(-41,-55){\vector(3,1){28}} \put(-79,-55){\vector(3,1){28}}

\put(-78,-61){$\gamma_{j-4}$}
\put(-50,-61){$\gamma_{j-3}$}
\put(-22,-61){$\gamma_{j-2}$}
\end{picture}
\vspace{14mm}

 %%% 19
 Now we consider all possible cycles for this graph and define the value of a cycle { {\large {\bf O}}} as the product of values of the edges included into the cycle multiplied by the number of the edges connecting the firstand last sites with a negative sign. When the numbers of the edges associated with the paths from the first site to the last one and back differ, we multiply by the maximum number of edges.  When the
cycle includes only one vertex, its value is defined as $E - \alpha_k$ where $k$ is the number of the vertex.  Then we introduce a cyclic term which is a set of cycles including all sites of the system. The value of the cyclic term is equal to the product of the cyclic
values of these cycles. The sum of all possible cyclic terms {\large {\bf O}}$_r$ gives us the cyclic value $\Theta(\Gamma_{ij})$ of the graph $\Gamma_{ij}. $ An expression for the matrix element $G_{ij}$ can be obtained in the form:
 \begin{equation} %f9
G_{ij}=\sum_{\sigma}\frac{P_{ij}^{\sigma}}{\Theta(\Gamma_{ij})}\ .
 \label{e9} 
          \end{equation}
 Here, the summation  is carried out over all possible
pathways along the chosen chain. The value of $P_{ij}^{\sigma}$ equals the product of all edges along the pathway. When the pathway includes all vertices of the graph $\Gamma_{ij}$ we have
       \begin{equation} %f10
P_{ij} = \prod_{k=i}^{j-1} \ \beta_k \ .
      \label{e10}
          \end{equation}
 When the pathway does not include some vertices, this produces different values for the corresponding $ P_{ij}$. For example, when the electron isbeing transferred from the site $l$ to the site $l+2$, passing the intermediate site $l+1$, the corresponding edge is equal to $(E - \alpha_{l+1}) \gamma_l $. Consequently, we obtain for the pathway which excludes this vertex:
   \begin{equation} %f11
P_{ij} =  
(E - \alpha_{l+1})\gamma_l  \prod_{k=i,\, k\neq l,l+1}^{j-1}
\ \beta_k\ .
          \label{e11}
      \end{equation}

 Following a similar way we can get explicit expressions for all terms in the numerator of Eq. (9). The expression (9) can be significantly simplified if we assume that (NNN)
couplings are small compared to (NN) couplings. Then we
obtain the following approximation:
   \begin{equation} %f12
\! \!\! G_{ij} = \frac{1}{\Theta_0(\Gamma_{ij})} \left \{ \!
\prod_{k=1}^{j-1}
\beta_k + \sum_{l=i}^{j-2} \! \gamma_l \left(E-\alpha_{l+1} \right ) \!\!\!\!\!\!\!
\prod_{k=i; k\neq l,l+1}^{j-1} \!  \right \} 
 \label{e13}
   \end{equation}
 where $\Theta_0(\Gamma_{ij}) $ is the cyclic value of the graph
calculated within the (NN) approximation.

 %%% 21

The Green's function matrix elements [Eq. (9)] have poles which can be detemined by solving the equations
     \begin{equation} %f13
\Theta_{ij} (\Gamma) =   0.
        \label{e14}
          \end{equation}
 These poles correspond to energy states of a given chain.  The set of these energy states characterizes the chosen chain (pathway) and different chains have different sets of energy states. When we take into account the broadening of  the bridge energy levels, the singularities of the Green's function matrix elements are converted to the peaks of finite height and width.
 
 To apply the described method of calculation of the matrix
elements $G_{ij}$ for a real macromolecule we have to determine which atomic orbitals have to be used to arrange a chain between the sites $''i''$ and $''j''.$  To choose suitable orbitals we can employ a rigorous quantum mechanical method of tunneling interatomic currents \cite{24} which was alredy successfully used to separate out electron transfer pathways in long-range ET processes \cite{25}. Data concerning the electronic structure of a considered molecule can be obtained within an extended Huckel \cite{25} or some other approximation \cite{26}. As a result we get the set of chains which represent the bridge, and we can use our result [Eq. (9)] in further analysis.

 It follows from the obtained results [Eqs. (7) and (9)] that the electron transmission function contains important information concerning intramolecular electron transfer. Under low temperatures and moderate electronic dephasing $T(E)$ can exhibit series of peaks. Their location is determined with contributions from donor/acceptor sites predominating in
the ET at a given interval of tunneling energy $ E,$ and with the energy spectrum of a chain of sites connecting them.  When  $E$ takes on a value close to $\epsilon_{k,l}$ or to one of the poles of the Green's function the corresponding term in Eq. (7) can surpass all remaining terms. Thus, for different values of $E $ different donor/acceptor sites and different pathways can predominate in the electron transfer through the bridge.

\begin{figure}[t]
\begin{center}
\includegraphics[width=6.0cm,height=6cm]{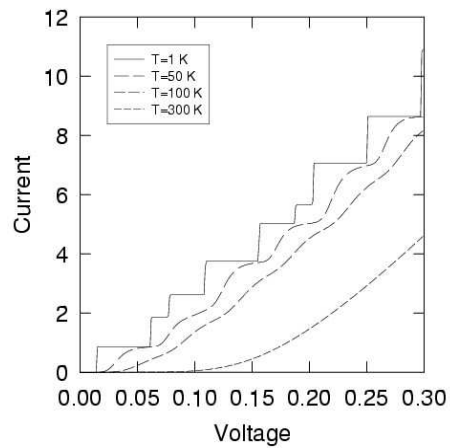}
\caption{
The calculated current (nA) -- voltage (V) characteristics for different temperatures.}   
\label{rateI}
\end{center}
\end{figure}

The structure of $T(E)$ can be revealed in the low temperature
current-voltage characteristics for the electronic transport through the molecular bridges  [Eq. (6)]. As a result of numerical simulations of electrical transport through porphyrin-nitrobenzene macromolecules it was shown \cite{27} that at low temperatures current-voltage curves can exhibit a steplike behavior which disappears as the temperature raises (see Fig. 1). This originates from a steplike character of Fermi distribution functions at low temperatures along with the ''comblike'' structure of $T(E).$ At a given voltage $V$ the difference of Fermi functions in Eq. (6) takes on nonzero values only in the interior of a certain energy range including $E_F.$ Therefore the magnitude of the current $I $ at a given voltage is determined by the contributions of peaks of $T(E)$  located in this energy range. When the applied voltage increases, this enhances the width of the relevant energy
interval. 
The tunneling current $ I $ abruptly changes when an extra peak of the electron transmission function appears there. Widths of the plateaus are equal to the distances between adjacent peaks of $ T(E) $ and magnitudes of sudden changes in the current correspond to the heights and shape of these peaks. At higher temperatures Fermi functions lose their steplike character and the plateous are washed out. Another reason for the structure of $ T(E) $ (and $ I-V$ curves) to be eroded is the electronic phase-breaking effect which arises due to stochastic fluctuatious of the ion potential. We conjecture, however, that at low temperatures the electronic dephasing effects could be reduced so that the structure of $ T(E) $ could be revealed.

On these grounds we believe that experiments on electronic transport through macromolecules at low temperatures $ (T \sim 1K )$ could provide an additional information upon characteristics of the electron transfer. Namely, comparing the structure of the electron transmission function reconstructed on the basis of experimental $ I-V $ curves with that obtained as a result of calculations using Eqs. (7) and (9), we can make conclusions concerning actual primary pathways of electrons, as well as sites of the donor/acceptor subsystems involved in the ET process at different values of the tunneling energy.  The model adopted in the present work provides us with the results suitable for a quantitative comparison with the results of proposed experiments even avoiding a proper and reliable calculation of the equilibrium Fermi energy of the bridge, and the effective voltage $V$. The lack of information about proper values of $ E_F $ and $V$ produces an uncertainty in the location of the origin at the ''voltage'' axis in the Fig. 1, so we cannot identify steps of the $I-V$ curves separately. Nevertheless, changes in the values of $E_F $ and
$ V $ do not influence the electron transmission function, therefore the structure of series of the peaks remains fixed. This enables us to identify some series of peaks analyzing sequences of widths of successive steps of the current-voltage curves. Such analysis also can give reasonable estimations for the $ E_F$ and $V$ for the chosen molecule \cite{28}.
\vspace{1mm}

{\it  Acknowledgments:}
I thank G.M. Zimbovsky for help with the manuscript.

%\end{multicols}


\begin{thebibliography}{99}
\bibitem{28} The equilibrium Fermi energy of the bridge can be expressed in terms of the bridge energy levels (Ref. \cite{16}). As for $V$, this quantity can be estimated from the width of the energy range where the relevant peaks of $T(E)$ are located.
\end{thebibliography}
\end{document}